## REVIEW

# Advances in InGaAs/InP single-photon detector systems for quantum communication

Jun Zhang[1,2], Mark A Itzler[3], Hugo Zbinden[4] and Jian-Wei Pan[1,2]

Single-photon detectors (SPDs) are the most sensitive instruments for light detection. In the near-infrared range, SPDs based on III–V compound semiconductor avalanche photodiodes have been extensively used during the past two decades for diverse applications due to their advantages in practicality including small size, low cost and easy operation. In the past decade, the rapid developments and increasing demands in quantum information science have served as key drivers to improve the device performance of single-photon avalanche diodes and to invent new avalanche quenching techniques. This Review aims to introduce the technology advances of InGaAs/InP single-photon detector systems in the telecom wavelengths and the relevant quantum communication applications, and particularly to highlight recent emerging techniques such as high-frequency gating at GHz rates and free-running operation using negative-feedback avalanche diodes. Future perspectives of both the devices and quenching techniques are summarized.

## INTRODUCTION

A single photon is the indivisible minimum energy unit of light, and therefore, detectors with the capability of single-photon detection are ultimate tools for weak light detection.[1,2] So far, single-photon detectors (SPDs) have been widely used in numerous applications such as quantum communication, quantum information processing, Lidar and photoluminescence. Most near-infrared SPDs can be sorted into three principal categories of devices: photomultiplier tubes, superconducting devices and semiconductor single-photon avalanche diodes (SPADs). Apart from these devices, there are also some new technologies for single-photon detection such as quantum-dot optically gated field-effect transistor[3] and quantum dot resonant tunneling diodes.[4] Up-conversion detectors[5–7] combining the nonlinear optical process of sum frequency generation and Si SPADs are still considered to belong to semiconductor devices.

Photomultiplier tubes are operated in high-vacuum tubes with high voltages between anodes and photocathodes.[8] A primary electron is produced in the photocathode material as a consequence of the photoelectric effect, and high gain results from a multiplication mechanism that creates secondary electrons. Photomultiplier tubes can have large active areas, but they suffer from low efficiency and high dark count rate.

Superconducting SPDs include superconducting nanowire single-photon detectors (SNSPDs),[9] transition edge sensors[10] and superconducting tunnel junctions.[11] In the SNSPD, a hotspot is created after the absorption of a single photon in superconducting nanowires, and subsequently, the superconducting current density increases due to the size expansion of the hotspot. Once the superconducting current density in the nanowires reaches the critical value, the nanowires are changed from the superconducting state to the normal resistance state. This transition generates a voltage signal of single-photon detection. The primary advantages of SNSPDs are low dark count rate, high photon count rate and very accurate time resolution. The detection efficiency was low (at the level of a few percent) for early generation devices, but recently, this parameter has been significantly improved through the efforts of the SNSPD community.[12] However, the cryogenic operating conditions required for SNSPDs limit their use for practical applications.

Currently the mainstream solution for single-photon detection in practical applications is the use of SPADs. In the literature covering photodetectors, one finds the terminologies of avalanche photodiode (APD) and SPAD. Normally, a device is referred to as an APD when it operates below the breakdown voltage in the linear-mode, for which the output photocurrent is linearly proportional to the input optical power. The term SPAD refers to a device operated in Geiger mode, for which biasing above the breakdown voltage can result in a self-sustaining avalanche in response to the absorption of just a single photon. For a SPAD based detector system, there are two crucial parts: the SPAD device[13–15] and the quenching electronics.[16–18] Therefore, the performance of SPD system depends both on the SPAD device itself and the quenching electronics as well.

[1]Hefei National Laboratory for Physical Sciences at the Microscale and Department of Modern Physics, University of Science and Technology of China, Hefei, Anhui 230026, China; [2]CAS Center for Excellence and Synergetic Innovation Center in Quantum Information and Quantum Physics, University of Science and Technology of China, Hefei, Anhui 230026, China; [3]Princeton Lightwave Inc., 2555 US Route 130 S., Cranbury, NJ 08540 USA and [4]Group of Applied Physics, University of Geneva, Geneva CH-1211, Switzerland
Correspondence: J Zhang, Hefei National Laboratory for Physical Sciences at the Microscale and Department of Modern Physics, University of Science and Technology of China, Hefei, Anhui 230026, China
E-mail: zhangjun@ustc.edu.cn

In the following sections, we will first introduce the basic semiconductor structure, device performance improvement, and Geiger mode operations for InGaAs/InP SPADs. The characteristics and the characterization methods relevant to InGaAs/InP SPADs are subsequently presented, and then we will focus on recent advances of quenching techniques, particularly in the regimes of low-frequency gating, high-frequency gating and free-running operation. The applications for diverse quantum communication protocols such as quantum key distribution (QKD),[19,20] quantum teleportation,[21] quantum secret sharing (QSS),[22–24] quantum secure direct communication[25–28] and counterfactual quantum cryptography[29] using InGaAs/InP SPADs will also be described briefly and representatively, and finally we conclude with a discussion of future perspectives on both SPAD devices and quenching techniques.

Relevant reviews concerning quantum cryptography,[30,31] quantum communication,[32,33] SPDs for quantum information applications,[1] single-photon sources and detectors,[2] solid-state SPDs[34] and avalanche photodiodes[35,36] could also be of significance to the reader as references.

## INGAAS/INP SPAD

For the single-photon detection in the near-infrared, group III–V heterostructure devices such as InGaAs/InP and InGaAs/InAlAs with separate absorption, grading, charge and multiplication structures[35,36] as shown in Figure 1 are the primary candidates. In these devices, an InGaAs ($In_{0.53}Ga_{0.47}As$) layer with a room-temperature band gap $E_g$ of 0.75 eV and a cutoff wavelength of around 1670 nm is used as the absorption material, while the lattice-matched InP layer or InAlAs layer is used as the multiplication material (Figure 1). The electric field in the multiplication layer is sufficiently high to provide the desired avalanche probability, while the electric field in the absorption layer is adequately low to minimize field-induced leakage currents.[13] The charge layer is designed to provide high electric field in the multiplication layer and low electric field in the absorption layer, while the grading layer avoids carrier accumulation in the heterojunction interface.[13] To improve SPAD performance, both the device structure design and device fabrication should be optimized specifically for single-photon detection.[37–39]

In Geiger mode, the reverse bias voltage of the SPAD ($V_b$) is larger than the breakdown voltage ($V_{br}$). When a photon is absorbed, an electron–hole pair of electrical carriers is created. One carrier is subsequently injected into the depletion zone of multiplication layer and may initiate a self-sustaining avalanche due to the impact ionization mechanism at high electric field (on the order of $10^5$ V cm$^{-1}$). The avalanche current reaches a macroscopic steady state within a buildup time on the order of a few hundred picoseconds.

The device structure of the InGaAs/InP SPAD illustrated in Figure 1 bears similarities to that of more mature 'linear-mode' APDs used at modest gains below their breakdown voltage. However, despite these structural similarities, the optimization of SPAD performance is significantly different from that of linear-mode APDs because these two device types are employed in dramatically different contexts.[13] Linear-mode APDs can provide sensitivity improvements in optical receivers (relative to more conventional receivers based on p–i–n photodiodes, which lack gain) as long as the noise of the APD is less than the noise of the amplifier which follows the APD in the receiver circuit. In general, linear-mode APDs only provide a sensitivity advantage for high-bandwidth (e.g., >1 GHz) receivers in which the necessarily broad frequency response leads to high amplifier noise. Therefore, linear-mode APD design emphasizes low excess noise and high bandwidth.[35,36]

In contrast, the role of SPADs is to provide an avalanche response that is sufficiently large to reliably detect the injection of a single photo-excited carrier into the multiplication region (Figure 1). This behavior is achieved by operating in Geiger mode (i.e., above $V_{br}$), and in this capacity, the SPAD is more appropriately described as a photon-activated switch with an essentially digital response that is noise-free, at least in the sense that the threshold for detecting avalanches can be set far beyond the level of any background circuit noise. The only noise in a SPAD originates in 'dark counts' induced by thermal or field-mediated mechanisms in the absence of input signal photons. While the average dark count level can be subtracted from the overall device output, the shot noise of these dark counts is unavoidable.

The different operating modes for linear-mode APDs and SPADs require optimization of distinct performance attributes. For instance, linear-mode operation benefits from a high gain-bandwidth product, and since gain-bandwidth product is nominally inversely proportional to the width of the multiplication region ($W_m$), linear-mode APD design tends to emphasize narrow $W_m$ of well under 1 μm. Conversely, gain-bandwidth product bears no direct impact on SPAD performance. Instead, wider multiplication regions can provide lower breakdown fields with a consequently smaller tunneling contribution to the dark counts, and so SPAD design tends to emphasize wider $W_m$ of well beyond 1 μm.[37,40] In a similar vein, linear-mode APDs benefit from lower excess noise, which can be provided by narrower multipliers, particularly when dead-space effects are prevalent.[41] However, excess noise is not directly relevant to SPAD performance, and design considerations related to excess noise are absent during SPAD design.

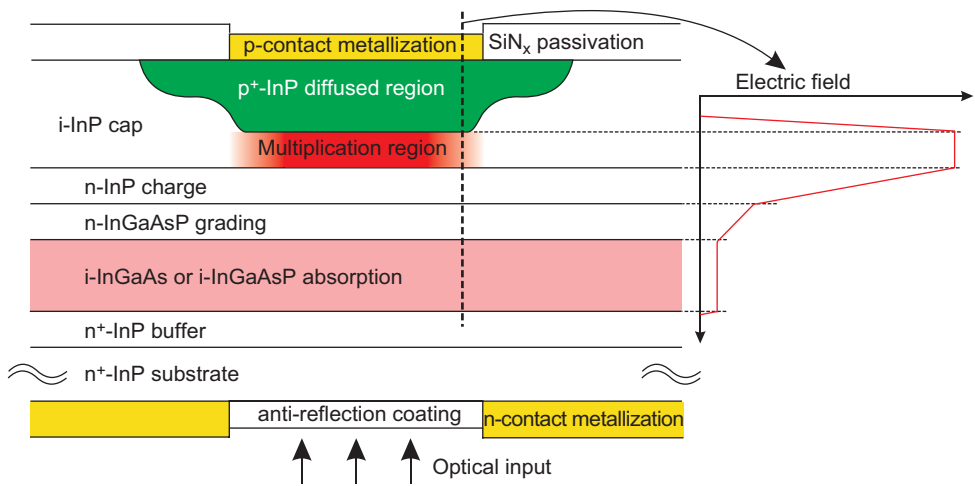

**Figure 1** The SAGCM structure of InGaAs/InP SPAD.[13] Figure reproduced: Ref. 13 © 2007, Taylor & Francis. SAGCM, separate absorption, grading, charge and multiplication; SPAD, single-photon avalanche diode.

From the perspective of underlying materials properties and fabrication technology in the InGaAsP material system, the sources of noise in linear-mode APDs and SPADs are also considerably different. One key consideration is that the dark current in state-of-the-art InGaAs/InP avalanche diode structures (including both linear-mode APDs and SPADs) is dominated by perimeter leakage mechanisms.[42] Although this perimeter leakage does not pass through the multiplication region and remains unmultiplied, for InGaAs/InP avalanche diodes at typical operating temperatures, it is still dramatically larger than the multiplied bulk leakage current. Therefore, the dark current performance of linear-mode APDs is dominated by perimeter leakage, and improvements in this device type will require wafer fabrication improvements such as better surface passivation techniques. For SPADs, however, because the perimeter leakage is not multiplied, it does not induce detection events, and only the multiplication of bulk dark carriers influences the dark counts. Therefore, the most profitable strategy for improvement beyond the state-of-the-art dark count in SPADs will include a focus on the bulk material properties of the base epitaxial wafers.

Beyond these considerations of underlying SPAD device design, overall device performance is critically dependent on the detection circuitry that follows the SPAD. An appropriate circuit, referred to as quenching electronics,[16] is necessary to rapidly suppress the avalanche by lowering the reverse bias down to $V_{br}$, to output the detection signal of the incident photon by discriminating the leading edge of the avalanche current and to restore the SPAD to its armed state to detect the next incoming photon. Note that rapid quenching also reduces afterpulsing as we will discuss below, therefore the quenching electronics plays a key role in a SPAD system. Quenching technologies include passive quenching, active quenching, gated quenching and hybrid quenching.[16]

In a passive quenching circuit,[43] a SPAD is connected with a reverse bias through a high-value ballast resistor (on the order of 100 kΩ). When avalanches occur, the voltage difference between the anode and the cathode of the SPAD decreases swiftly due to the voltage drop across the resistor. Once the voltage difference is lowered sufficiently close to $V_{br}$, avalanches will spontaneously quench. For a well-designed SPAD and associated quenching circuit, the quenching time, defined as the time duration from avalanche occurrence to avalanche termination, is on the order of 1 ns. Passive quenching is well suited for asynchronous single-photon detection (free-running mode). However, this technique suffers from long recovery times on the order of 100 ns due to the large time constant of resistance capacitance (RC).

To solve the problem of slow recovery in passive quenching, active quenching is implemented[44,45] by using fast discrimination electronics to sense the leading edges of avalanches. The output signals of the discrimination electronics are used to switch the SPAD bias below $V_{br}$, and the device is maintained in this disarmed state for a certain time period called the hold-off time or deadtime. Following the hold-off time, the bias is actively switched back to the initial armed state. In this scheme, both the quenching time and the recovery time can be a few nanoseconds or less using commercially available electronics components.

Gated quenching[46] is widely used for applications requiring synchronous single-photon detection such as QKD.[30] In such a scheme, gate pulses with a repetition frequency of $f_g$, voltage amplitude of $V_g$ and time duration of $\tau_g$ (with a resulting duty cycle of $\tau_g f_g$) are coupled to a SPAD. The SPAD is working in Geiger mode only when the pulses are gated on. However, avalanche signals generated during the gate pulses are superposed with parasitic derivative signals that result from capacitive responses of the SPAD to the gate pulses. The key technical challenge in gated quenching is to extract avalanche signals from the background capacitive responses.

Each quenching technique has its own advantages and disadvantages. Therefore, hybrid quenching schemes combining the above techniques are sometimes applied. For instance, the scheme of passive quenching and active reset, which will be introduced in the later section, can efficiently shorten the recovery time. In the gated quenching scheme, for long gate width, the use of active quenching instead of gated quenching can significantly reduce the quenching time.

## CHARACTERISTICS AND CHARACTERIZATION OF SPAD

For a SPAD-based SPD system, there are quite a few parameters that are important for performance evaluation, and optimization of one parameter often involves performance tradeoffs with other parameters. In this section, we first introduce the definitions and the mechanisms of these parameters, and then describe the experimental characterization.

The first important parameter is (single) photon detection efficiency (PDE),[40] defined as the probability that the detector system produces a desired output signal in response to the arrival of an incident photon. From the view of SPAD structure, PDE is determined by PDE = $\eta_{coup} \times \eta_{abs} \times \eta_{inj} \times \eta_{ava}$, where $\eta_{coup}$ is the coupling efficiency of SPAD, $\eta_{abs}$ is the absorption efficiency or (internal) quantum efficiency in the absorption layer of SPAD, $\eta_{inj}$ is the collection efficiency of the photo-excited carriers injected from the absorption layer to the multiplication layer and $\eta_{ava}$ is the probability of a detectable avalanche given the successful injection of a carrier into the multiplication layer. $\eta_{coup}$ depends on multiple factors such as insertion loss, surface reflectance and active area of device. $\eta_{abs}$ is calculated by $\eta_{abs} = 1 - e^{-\alpha d}$, where $\alpha$ is the absorption coefficient and $d$ is the absorption depth. For InGaAs, $\alpha$ is around 7500 cm$^{-1}$ at 1550 nm illumination. Given $d = 1.5$ μm, $\eta_{abs}$ is 0.68. $\eta_{inj}$ and $\eta_{ava}$ have a sensitive dependence on the electric field, which is determined by the excess bias ($V_{ex}$) defined as $V_{ex} = V_b - V_{br}$.

Dark count rate (DCR)[47] is used to characterize the noise performance of the detector system. DCR is defined as the normalized count rate in the absence of illumination. DCR depends on the conditions of temperature ($T$) and $V_{ex}$. Dark counts originate from the mechanisms of thermal excitation, tunneling excitation or trap-assisted tunneling excitation. At sufficiently high operation temperatures, thermal excitation will be the dominant contribution to DCR, while at low temperatures or high electric fields (large $V_{ex}$), tunneling excitation will dominate the contributions to DCR.[13] Although DCR is analogous to APD dark current in the linear-mode—e.g., the shot noise of both of these phenomenon plays a comparable role in their overall noise performance—DCR is generally not correlated with the device dark current measured in the linear-mode. This is due to the fact that dark current in the linear-mode is usually dominated by perimeter leakage currents that do not flow through the multiplication region and therefore do not result in Geiger mode avalanches.

Afterpulse probability ($P_{ap}$) is another important parameter of SPAD. During an avalanche, some carriers are trapped by defects and impurities in the multiplication layer. Subsequently these carriers are released and can initiate new undesired avalanches called afterpulses.[48,49] $P_{ap}$ is defined as the probability of producing afterpulsing counts due to the previous photon detection during a time period. Reducing $P_{ap}$ to an appropriately low level is crucial for most applications. Improving the crystal quality of multiplication material can effectively suppress the afterpulsing effect, but advances in fundamental material quality such as significant reduction in defect density are likely to take many years given the current relative maturity of the InGaAsP materials system. Alternatively, reducing the quantity of charge carriers during the avalanche process or shortening the lifetime of trapped carriers can also decrease $P_{ap}$. For a detector system, $P_{ap}$ is related to multiple conditions, which can be roughly modeled as $P_{ap} \propto (C_d + C_p) \times \int_0^\delta V_{ex}(t)dt \times e^{-\tau_d/\tau}$, where $C_d$ is the diode capacitance, $C_p$ is the parasitic capacitance of circuit including the lead capacitance of device, $\delta$ is the avalanche duration time, $\tau_d$ is the hold-off time and $\tau$ is the lifetime of detrapping carriers. From the above equation, the approaches to reduce $P_{ap}$ include: (1) minimizing $C_p$; (2) limiting $\delta$; (3) lowering $V_{ex}$; (4) increasing $\tau_d$; and (5) decreasing $\tau$ by increasing the operation temperature. However, these approaches also have consequent disadvantages. For instance, approach (1) and approach (2) can effectively reduce avalanche charge quantity, but the technical challenge is to extract weak avalanches from background noise. Approach (3) decreases PDE. Approach (4) limits the maximum count rate. Approach (5) increases DCR. Therefore, obtaining the desired afterpulsing performance can often force compromises with other SPAD performance parameters.

Timing jitter (time resolution) is usually defined as the total time uncertainty between incident photons and corresponding electrical signal outputs, which includes the contributions of the SPAD itself and the quenching electronics as well. The intrinsic timing jitter of the SPAD device strongly depends on $V_{ex}$. With large $V_{ex}$, high electric field greatly shortens avalanche build-up time and also its time uncertainty.

Maximum count rate is determined by the capability of the SPAD in the limit of saturated photon detections, which can be approximated by the reciprocal of $\tau_d$. The ratio of the maximum count rate and the DCR determines the dynamic range of the SPAD.

Photon number resolution (PNR) is important for specific applications such as quantum computation and quantum optics. With conventional techniques of gating mode and free-running mode, the SPAD is believed not to have the capability of PNR since avalanches always reach the saturation stage due to relatively long quenching time. Operated in such a fashion, the SPAD can only resolve between zero photons and non-zero photons. Using a SPAD array (multi-pixels) or time-multiplexing scheme[50] can implement PNR with conventional quenching techniques. Recently, experiments have shown that with ultra-short quenching time as described in later section, for instance, using the technique of high-speed gating, SPAD can also exhibit PNR capability.

The methods to characterize these parameters include the single-photon scheme[51,52] and photon pair scheme.[53] In the single-photon scheme,[51,52] a calibrated optical power meter with high accuracy and low uncertainty is used to measure the power of a pulsed laser, and a calibrated attenuator is used to highly attenuate the laser power down to the single-photon level. Therefore, the calibration for these optical instruments themselves is crucial in such a method. In the photon pair scheme,[53] correlated photon pairs are generated via a nonlinear optical technique such as spontaneous parametric down-conversion, in which signal and idler photons are sent to two detectors for calibrating and triggering, respectively. The output signals of detectors are further processed by a coincidence counter. In this method, the SPAD parameters, particularly PDE, can be precisely measured without requiring a calibrated power meter. Nevertheless, coupling efficiency in each channel is an important factor to be carefully considered.

A typical experimental set-up of the single-photon calibration method is shown in Figure 2. The SPAD is working in the gated mode with low repetition frequency ($f_g$), e.g., 10 kHz. A signal generator outputs original gates and synchronized signals with the same frequency to drive a laser diode. Short optical pulses emitted from the laser diode are divided by a beam splitter. One port of the beam splitter is monitored by a power meter and the other one is connected with a variable attenuator. The power of pulses after the attenuation reaches quasi-single-photon level. The detection signals produced in the quenching electronics are finally connected with a counter. To well illustrate the calibration and the tradeoffs of the SPAD parameters, we construct an experimental single-photon test bench according to Figure 2 and characterize an InGaAs/InP SPAD as an example. The typical experimental results are plot in Figure 3.

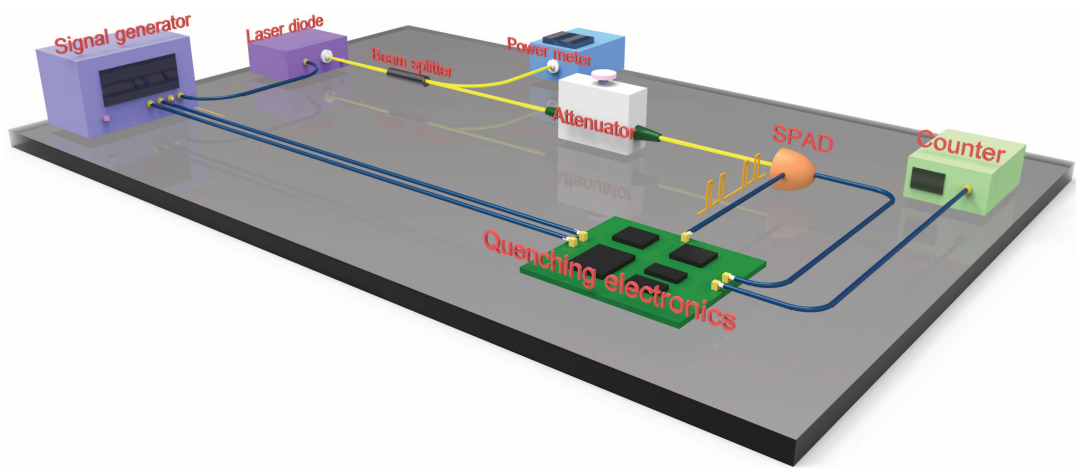

**Figure 2** Single-photon test bench for SPAD performance calibration. SPAD, single-photon avalanche diode.

Given that measured count rates are $C_{on}$ and $C_{off}$ with and without laser illumination, respectively, the normalized DCR can be calculated as $\text{DCR} = \frac{C_{off}}{f_g t_w} \times 10^9$, where $t_w$ is the effective gating width (in ns). Considering the Poisson distribution of incident photons, PDE can be calculated as $\text{PDE} = \frac{1}{\mu} \ln \frac{1 - C_{off}/f_g}{1 - C_{on}/f_g}$, where $\mu$ is the mean photon number per optical pulse.

The characterization of PDE and DCR as $V_{ex}$ linearly increases is shown in Figure 3a. Apparently, PDE is a linearly function and DCR is an exponential function of $V_{ex}$. Therefore, the relationship between DCR and PDE is exponential.

To precisely characterize $P_{ap}$, double-gate method[51] is widely used. The first gate is for photon detection while the second gate with tunable delay relative to the first gate is for measuring afterpulse counts. In such a way, by varying the delay between the two gates the afterpulsing distribution can be clearly plotted after the subtraction of DCR contribution.

The experimental results of the afterpulsing distribution at different temperatures are plot in Figure 3b, in which the normalized afterpulse probabilities ($P_{ap}/t_w$) exponentially decay in time. The gap between the two lines in Figure 3b clearly shows that increasing temperature can effectively suppress the afterpulsing effect.

Timing jitter is usually measured using an instrument capable of time-correlated single-photon counting or a common time-to-digital converter. The detection signal in the quenching electronics and the synchronized output from the signal generator can be used as 'Start' and 'Stop' of the timing measurement instrument, respectively. After eliminating the intrinsic jitter due to the signal generator, the laser, and the timing measurement system, the timing jitter of the whole detector system can be obtained.

## LOW-FREQUENCY GATING

Gated mode is a simple and effective approach to suppress DCR and afterpulses for synchronous single-photon detection.[46] When the electronic signals of the gates are coupled to a SPAD, derivative signals are created due to the capacitive response of the SPAD, and avalanche signals are superimposed on the derivative signals. Suppressing the derivative signals to effectively extract avalanche signals is the key task in the gated quenching electronics. The amplitudes of derivative signals depend on the rise (and fall) time of gates, gate amplitudes and quenching circuits. The amplitudes of avalanche signals depend on $V_{ex}$ and $t_w$. $t_w$ is the most important parameter in the gated quenching scheme. If $t_w$ cannot be too short, e.g. 1 ns or less, the afterpulsing contribution is still considerably high. To suppress the afterpulsing effect, a long hold-off time at the level of microseconds is necessary, which substantially limits the gating frequency. Using conventional gating techniques, the maximum frequency is limited to a few tens of MHz.

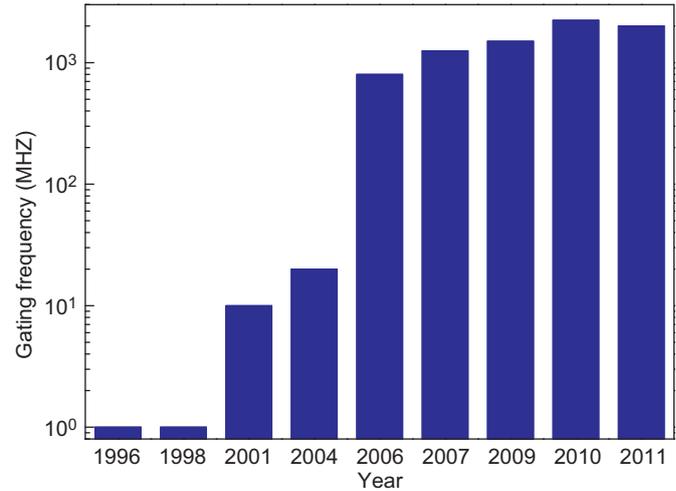

**Figure 4** The evolution of gating frequency for InGaAs/InP SPADs. All the data are taken from the references. SPAD, single-photon avalanche diode.

The evolution of gating frequency is shown in Figure 4. Dramatic increases in gating frequency occurred during the dozen years between 1998 and 2010. In this Review, we define 100 MHz as a critical point, which means that gating techniques with frequency below (above) 100 MHz are categorized as low-frequency (high-frequency) gating.

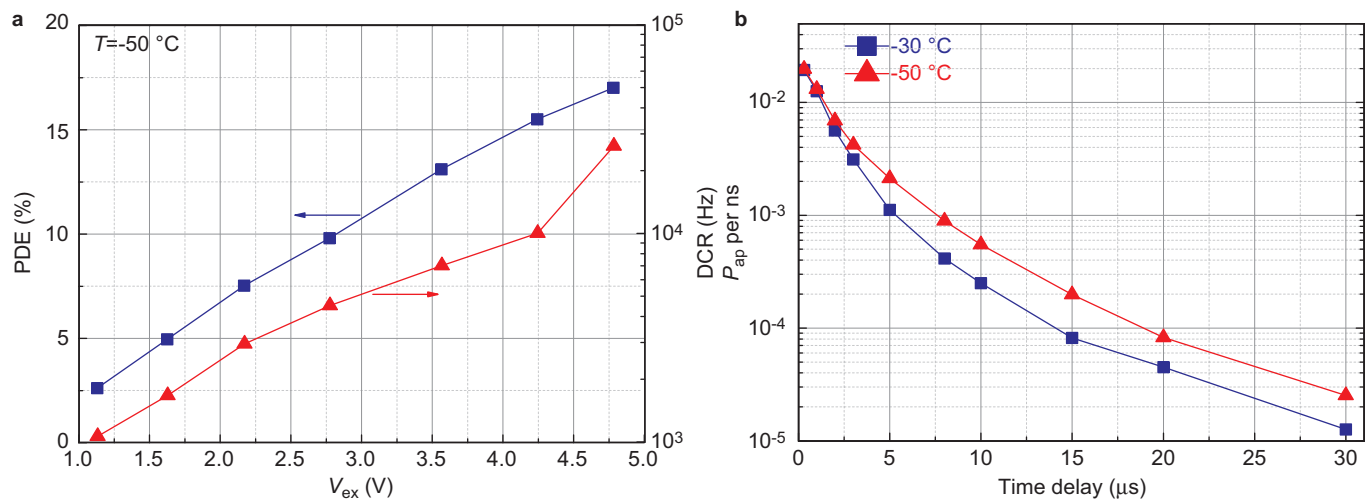

**Figure 3** Experimental characterization of an InGaAs/InP SPAD. (**a**) PDE and DCR versus $V_{ex}$ at −50 °C. (**b**) Normalized parameter of $P_{ap}$ per ns as a function of time delay between two gates with PDE=10% at −30 °C and −50 °C, respectively. In the experiment, $f_g$=10 kHz, $\mu$=1 and $t_w$=50 ns. DCR, dark count rate; PDE, photon detection efficiency; SPAD, single-photon avalanche diode.

In this section, we describe the development and evolution of low-frequency gating, and introduce some representative gating techniques among numerous works[46,51,54–75] in the past, which could be good references to invent new quenching techniques.

The coincidence method[64] is a standard technique for avalanche extraction in low-frequency gating. Electronic gate signals, as shown in Figure 5b(1), are alternating current (AC) coupled to the cathode of SPAD. Output signals with the superposition of avalanches and derivative signals at the anode of SPAD (Figure 5b(2)) are discriminated by a comparator. The comparator outputs (Figure 5b(3)) and the auxiliary signals synchronized to the gates (Figure 5b(4)) are inputs to an AND logic gate, whose outputs (Figure 5b(5)) are effective avalanche events. The timing of the auxiliary signals is precisely controlled to avoid the coincidence with the discrimination outputs of the derivative signals. The coincidence method can be easily implemented using analog and digital circuits. However, there are still some drawbacks in such a scheme. For instance, to avoid false detections due to electronic noise, the threshold of the comparator is necessarily increased. This may result in small avalanches not being detected, e.g., avalanches that have less time to build up because they occur near the end of gates. If the derivative signals are well suppressed, the amplitude ratio of avalanches to background signals can be effectively improved and hence, the comparator threshold can be further lowered.

Common-mode cancellation is the basic idea behind suppressing capacitive responses, and there are diverse approaches to cancelling the derivative signals. In 2000, Bethune and Risk[57] reported a transient cancellation technique using radio frequency (RF) delay lines, and based on this detector, they implemented an autocompensating quantum cryptography system.[57,59] Details about the detector can be found in Ref. 63.

The RF delay line scheme and its timing diagram are shown in Figure 6. Electronic pulses (Figure 6b(1)) are AC coupled to the cathode of the SPAD via a directional coupler and are also connected with an open-circuit cable whose length is $L$. The non-inverted reflection at the end of the open-circuit cable forms an additional gate pulse (Figure 6b(2)) delayed by a time interval of $2L/v$ relative to the original pulse, where $v$ is the propagation speed of electromagnetic waves in the cable. The anode of the SPAD is connected to another cable whose length is the same as the upper cable, but a short-circuit termination of this second delay cable results in an inverted reflection. The transient signal at the anode is the superposition of the SPAD response signal due to the initial gate pulse and its non-inverted reflection from the upper cable along with its inverted reflection that is delayed by $2L/v$. As a result of this superposition, the parasitic derivative signals from the two reflections cancel each other, and with appropriate timing of photon arrivals the avalanche signal clearly stands out (Figure 6b(3)). After passing the controlling gate (Figure 6b(4)) and amplifier, the amplified avalanche signals (Figure 4b(5)) then can be easily discriminated. With effective cancellation, the discrimination threshold is substantially lowered, which allows for further reduction of the gating width. As a consequence of shorter gate, the afterpulsing performance of the SPAD is improved. The primary drawback of this approach is that the intrinsic delay due to the cable reflections severely limits the maximum gating frequency.

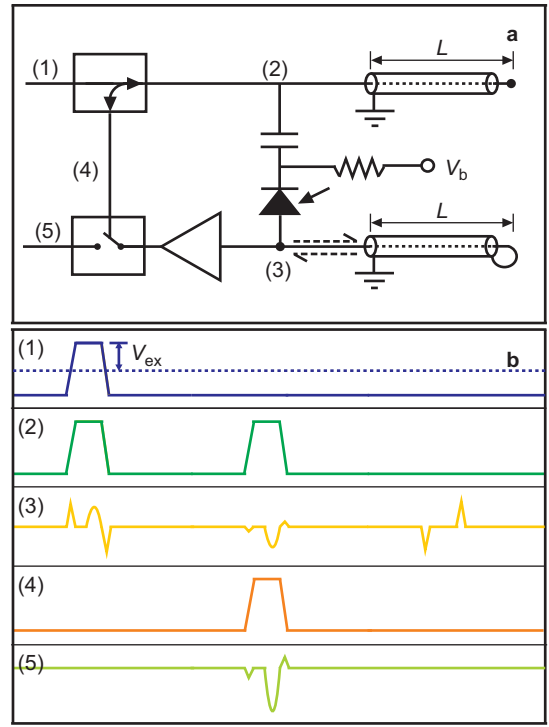

**Figure 6** Scheme (**a**) and timing diagram (**b**) of RF delay line cancellation method for low-frequency gating. $V_b$, bias voltage; $V_{ex}$, excess bias; $L$, cable length.

In 2002, Tomita et al.[60] implemented a two-channel detector system using the double-SPAD technique. Two SPADs are operated as shown in Figure 7. These two SPADs have very similar semiconductor parameters such as diode capacitance and excess bias-efficiency relationship.

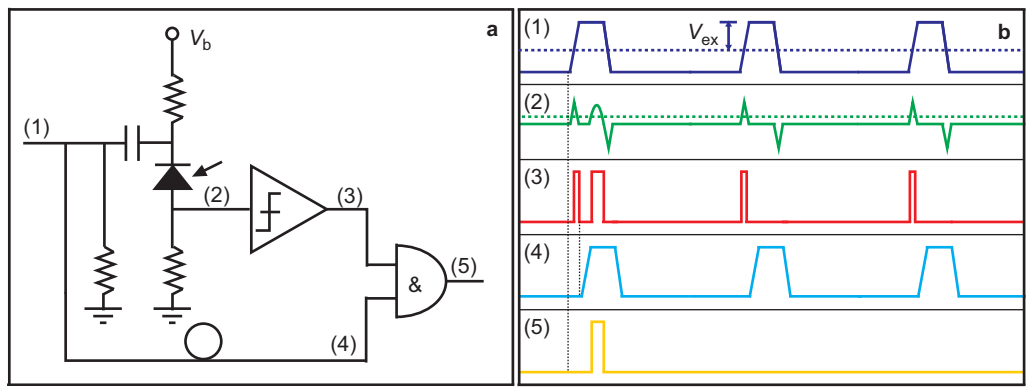

**Figure 5** Scheme (**a**) and timing diagram (**b**) of coincidence method for avalanche extraction in low-frequency gating. $V_b$: bias voltage. $V_{ex}$: excess bias. The signal amplitudes in this and the following figures are not to scale.

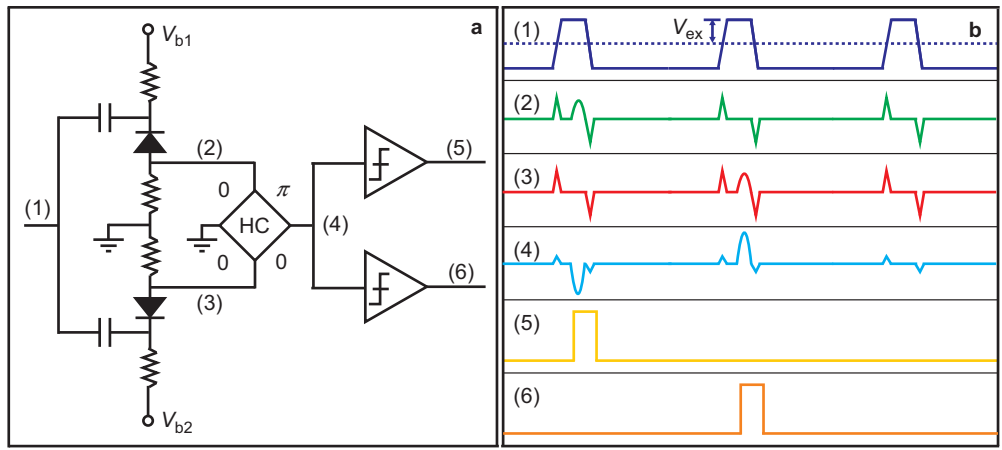

**Figure 7** Scheme (**a**) and timing diagram (**b**) of double-SPAD cancellation method for low-frequency gating. $V_{b1}$, $V_{b2}$, bias voltages; $V_{ex}$, excess bias. HC, hybrid coupler; SPAD, single-photon avalanche diode.

Electronic pulses (Figure 7b(1)) are AC coupled to the cathodes of both the SPADs in parallel. The anodes of SPADs are connected with a 180° hybrid coupler (HC). Due to the similar capacitive responses, the derivative signal shapes at the anode of the upper SPAD (Figure 7b(2)) and that at the anode of the lower SPAD (Figure 7b(3)) are almost the same, given that avalanches of the two SPADs are created at different gates. After passing the HC, the derivative signals cancel each other. The avalanche of the upper SPAD is reversed and the avalanche of the lower SPAD remains the same polarity, as shown in Figure 7b(4). The avalanches are then discriminated by two comparators with negative (Figure 7b(5)) and positive (Figure 7b(6)) thresholds, whose outputs indicate detection clicks at the upper and the lower SPADs, respectively.

This double-SPAD technique also has some drawbacks. The primary drawback is avalanche cancellation. When avalanches of the two SPADs are created at the same gate, the two avalanche signals will cancel each other due to the HC. Therefore, the two detectors cannot output detection clicks at the same time. For some specific applications, this drawback is still acceptable. When a single photon arrives at two-port devices such as beam splitter and polarizing beam splitter, the photon can only be detected by one of the SPADs. In a QKD system, this technique is well suited for the receiver integration. In addition, to maximize the suppression ratio for this scheme, selecting two SPADs with very similar parameters is necessary, which may be achieved with efforts of device screening in practice.

There are also some variant techniques to avoid the above drawbacks. For instance, using a common diode to replace one of the SPADs is a practical solution to avoid both the problems of avalanche cancellation and device screening at the cost of a potential decrease in the suppression ratio. The HC as shown in Figure 7 is the key component to eliminate the derivative signals. Besides the HC, there are some common devices that can implement the function of signal suppression such as common-mode choke coils, radiofrequency transformers and differential amplifiers. In low-frequency gating, these common devices may be used for practical implementations.

## HIGH-FREQUENCY GATING

Increasing the gating frequency for SPADs is critical for applications requiring high count rate. Actually, the invention of high-frequency gating techniques was originally driven by high-rate QKD applications. For a point-to-point QKD system, the raw key rate ($R_{raw}$) is roughly calculated as[30] $R_{raw} \propto \frac{1}{2} f \mu t \eta$, where $f$ is the system clock frequency that is usually the same as $f_g$, $t$ is the channel transmission, and $\eta$ is detection efficiency. Given a QKD system operating at a certain distance, $t$ and $\eta$ are fixed. Also, for security consideration, $\mu$ cannot be simply increased. Therefore, increasing $f$ is the only way to achieve higher bit rate.[30]

As explained in the above sections, the primary obstacle of high-frequency gating is the afterpulsing effect. One of the most effective approaches for afterpulsing suppression is reducing $t_w$. Hence the key technology challenge of high-frequency gating is weak avalanche extraction among strong stray signal when $t_w$ is ultrashort. In 2006, Namekata et al.[76] first reported a high-frequency gating detector with a clock rate of 800 MHz. The technique is called sine wave gating (SWG), as shown in Figure 8.

In this scheme, sine waves (Figure 8b(1)) with peak-peak amplitude of $V_{pp}$ are used as gates. Since the frequency spectrum of an ideal sine wave is pure, the response signals are only composed of sine waves with the fundamental frequency and higher order harmonics. Avalanches are superimposed over the response signals (Figure 8b(2)). These response signals can be easily eliminated by cascaded band-stop filters at center frequencies of $f_g$, $2f_g$, $3f_g$, etc. After the process of filtering-amplification-filtering, weak avalanches can be finally extracted (Figure 8b(3)) and discriminated (Figure 8b(4)). Low-pass filters are normally used before the comparator in order to smooth the analog signals of amplified avalanches.[76]

When $f_g$ is at the level of GHz, $t_w$ is generally very short and this parameter may be further reduced by tuning $V_{pp}$. Ideally, $t_w$ can be calculated as $t_w = \left(1 - \frac{2}{\pi} \arcsin \frac{2V_{ex}}{V_{pp}}\right) / f_g$. Empirically, in a GHz SWG detector system, $t_w$ can be as short as around 200 ps, which drastically suppresses the afterpulsing effect and hence greatly improves the count rate. The maximum count rate of the SWG scheme can in principle reach the same value as $f_g$. However, due to remaining afterpulsing contributions, a small hold-off time of tens of ns may still be necessary, which limits the maximum count rate to the range of tens of MHz. This is higher than low-frequency gating detectors, often by several orders of magnitude. Since the conventional hold-off time method cannot be directly applied to the SWG scheme, 'count-off time' is normally utilized instead.[77]

In 2007, Yuan et al.[78] implemented a new high-frequency gating technique called self-differencing. As shown in Figure 9, square waves with a clock rate of GHz (Figure 9b(1)) are coupled to the cathode of the SPAD. The response signals at the anode are first split by a power divider, and signals at one of output ports are delayed by one period. The two signals (Figure 9b(2) and Figure 9b(3)) are then subtracted from each other. As a result, the derivative signals can be effectively eliminated. After amplification, avalanches (Figure 9b(4)) with positive and negative parts can be easily discriminated. The self-differencing method is similar to the double-SPAD technique introduced in the previous section. For practical implementation, the subtraction circuit can be achieved using devices such as RF transformers. Similarly, the drawback of avalanche cancellation also exists when avalanches occur in adjacent gates. Therefore, the maximum ideal count rate is only $f_g/2$.

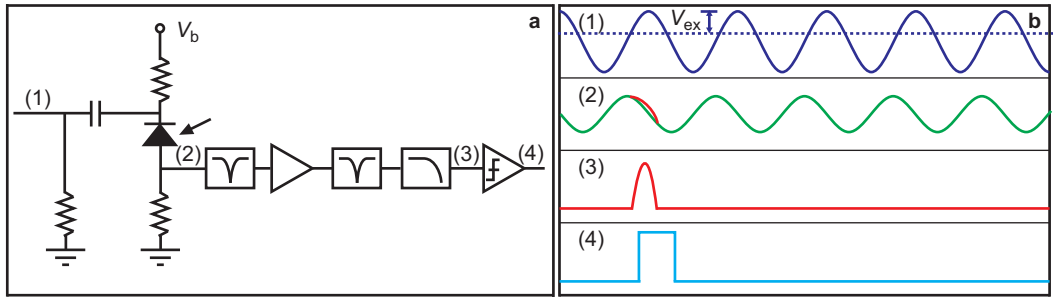

**Figure 8** Scheme (**a**) and timing diagram (**b**) of sine wave gating for high-frequency gating. $V_b$, bias voltage; $V_{ex}$, excess bias.

Presently sine wave gating and self-differencing are the two fundamental techniques used for high-frequency single-photon detection. Each technique has its own advantages and disadvantages in practice. So far, many groups have already implemented high-frequency detector systems using these two techniques or improved schemes,[76–94] and the clock rate has been increased to over 2 GHz.

In 2009, the Geneva group reported a practical solution for high-frequency gating by combining sine wave gating and self-differencing together.[77] Sine waves were applied as gates, and filters were used to partly reduce response signals while a self-differencing circuit was used to finally eliminate response signals. In such a way, the suppression requirement for the filtering circuit and the self-differencing circuit was moderate compared to that of using each technique individually.[77] Wu et al.[95] demonstrated an optical self-differencing method. The response signals of the SPAD were used to drive a laser diode after amplification, and then the optical pulses emitted from the laser diode with the same shapes as the response signals were divided by an optical beam splitter. One output port of the beam splitter was delayed by one period, and two channels of optical pulses were coupled to two balanced photodiodes for cancellation, respectively.

In 2010, the Geneva group presented a SWG detector with a clock rate of 2.23 GHz.[82] This frequency is the fastest reported so far and approaches the bandwidth limit of current commercial InGaAs/InP SPADs. Applying such detectors to QKD applications, simulations show that the detector performance is already comparable to a common SNSPD, and the maximum communication distance can reach around 200 km.[82] Chen et al.[84] invented an improved gating technique called double self-differencing. This cascaded self-differencing circuit could more effectively reduce response signals and thus further improve the SNR of avalanches.

In 2012, Liang and co-workers[88] first developed a stand-alone instrument consisting of a fully integrated SPD system with 1.25 GHz sine wave gating. The 2U rack instrument included diverse functionalities such as precise controls for temperature, bias, amplitude, comparator threshold, delay, and friendly user-interface and relevant auxiliary hardware interfaces. Walenta et al.[89] also implemented a 1.25 GHz SWG detector module using only low-pass filters, which was well suited for QKD system integration. Similar work has also been done by Liang et al.[87] and based on the 1 GHz gating detector, they demonstrated a laser ranging experiment.[85]

Apart from sine waves and square waves, in 2013, Zhang et al.[93] demonstrated a high-frequency gating detector using Gaussian pulses. Given time-dependent Gaussian pulses $G(t) = \frac{1}{\sqrt{2\pi}\sigma} e^{-t^2/2\sigma^2}$, where $\sigma$ is related to pulse width, its derivative signals can be written as $\frac{dG(t)}{dt} = \frac{-t}{\sqrt{2\pi}\sigma^3} e^{-t^2/2\sigma^2}$. The exponential term in the derivative is the same as that in $G(t)$. The falling edge shape of the bipolar response signals is similar to that of the original Gaussian pulses. If this falling edge is combined with the rising edge of a referencing attenuated and phase-matched Gaussian pulses which are synchronized to the original one, the falling edge can be cancelled and a steady plateau is formed. Avalanches are superimposed onto the plateau and hence, can be easily discriminated.

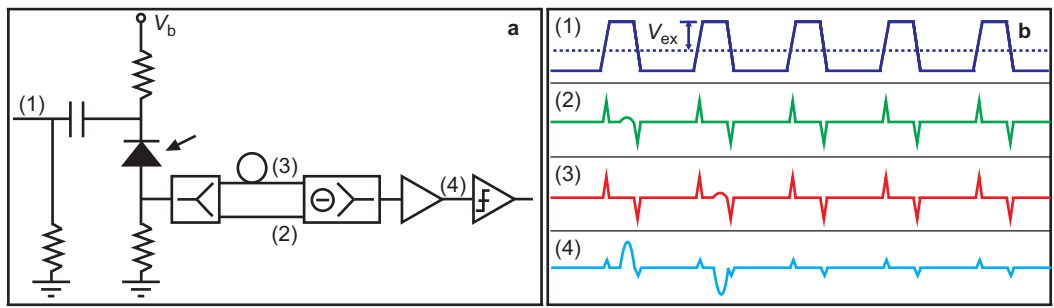

**Figure 9** Scheme (**a**) and timing diagram (**b**) of self-differencing for high-frequency gating. $V_b$, bias voltage; $V_{ex}$, excess bias.

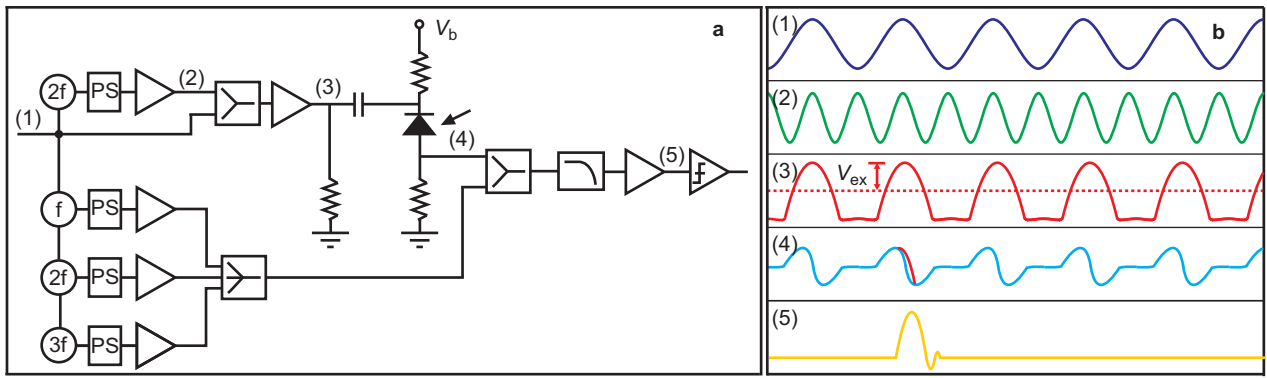

**Figure 10** Scheme (**a**) and timing diagram (**b**) of harmonics subtraction for high-frequency gating. $V_b$, bias voltage; $V_{ex}$, excess bias. PS, phase shifter.

Recently, Restelli *et al.*[94] used harmonics subtraction instead of band-stop filters for SWG, which could bring better afterpulsing performance compared with the standard filtering method. The harmonics subtraction method is shown in Figure 10. A sine wave (Figure 10b(1)) and its second harmonic (Figure 10b(2)) are combined and then amplified up to 20 V peak-to-peak (Figure 10b(3)). Although the amplified gates are distorted, the rising and falling edges are steeper than those of the fundamental sine wave. High amplitude and steep slope can further shorten the gating width compared with the standard SWG scheme. The first, second and third harmonics are synthesized with phase and amplitude control and then combined with the response signals (Figure 10b(4)) at a power divider. As a result, frequency components due to the harmonics are removed so that remaining avalanches are easily extracted after passing through a low-pass filter and a low-noise amplifier (Figure 10b(5)).

SPADs are traditionally believed to output only click or non-click for incident photons, but not to resolve photon numbers. However, with the high-frequency gating techniques SPADs may have the capability of PNR. Owing to ultra-short avalanche duration, weak avalanches are always in the sub-saturation stage so that different numbers of photon-generated carriers can result in different avalanche amplitudes. Experiments of PNR using high-frequency gating SPADs have been demonstrated.[95–97] However, the incident photon numbers, cannot be effectively resolved, due to the low detection efficiency.

In all the above high-frequency gating schemes, InGaAs/InP SPADs have been used. However, these techniques can be widely applied for other SPADs of different materials. For instance, photon number resolving[97] and high efficiency[98] Silicon SPADs using high-frequency gating techniques have been reported.

## FREE-RUNNING OPERATIONS

Passive quenching[16] is the fundamental approach for free-running single-photon detection,[43] and this scheme was initially demonstrated about two decades ago. However, there are still two technical problems to be solved. One is long recovery time and the associated baseline shift due to avalanche pileups during recovery. The other is large afterpulse probability. The recovery problem could be overcome using a resetting circuit in practice. Thus, the afterpulsing problem remains the key challenge in implementing free-running detectors. Simply increasing the hold-off time severely limits the count rate. Some groups reported different feasible methods to reduce the afterpulsing contribution. Warburton *et al.*[99] implemented a free-running InGaAs/InP SPAD through carefully optimizing the operation conditions such as lowering excess bias and increasing temperature. The Virginia group implemented a sophisticated method for the afterpulsing reduction,[100–102] i.e., removing the package of device and connecting the contacts of the SPAD and the quenching circuit by chip-to-chip wire bonding. In such a way, the stray capacitance was minimized.

Apart from passive quenching, free-running SPADs using active quenching have also been reported. In such cases, reducing the quenching time was crucial for the afterpulsing suppression. The Geneva group implemented a free-running detector based on an active quenching ASIC to minimize the parasitic capacitance of the electronics.[103]

High-frequency gating can also be used to mimic free-running operation, and relevant experiments for applications have been demonstrated.[85,104] The advantages of this method are low afterpulsing probability and high count rate, while the disadvantage is photon loss due to the duty cycle. For instance, given a gating frequency of 1 GHz and a gating width of 200 ps, the equivalent detection efficiency in the free-running mode is only 1/5 of that in the gating mode.

In this section, we will focus on the recent progress of free-running detectors including passive quenching and active reset (PQAR) and negative feedback avalanche diodes (NFADs).

Figure 11 shows a typical PQAR scheme implemented by the Virginia group.[101] The large resistor in the standard passive quenching schemes is replaced by a high-frequency GaAs FET. The FET is appropriately biased to hold the off-state. In the off-state, the FET has high impedance so that avalanches can be quenched passively. The avalanche signals are direct current coupled to a low-noise amplifier at the cathode of the SPAD. The discrimination outputs of the amplified avalanches are used to drive a pulse generator, which creates reset signals after the period of the hold-off time. The reset signals are AC coupled to the gate terminal of the FET to activate the FET into the on-state. In the on-state, the FET has low impedance so that the voltage at the anode of the SPAD can be rapidly recharged to the initial value ($V_b$). This PQAR scheme offers good dynamic range performance.

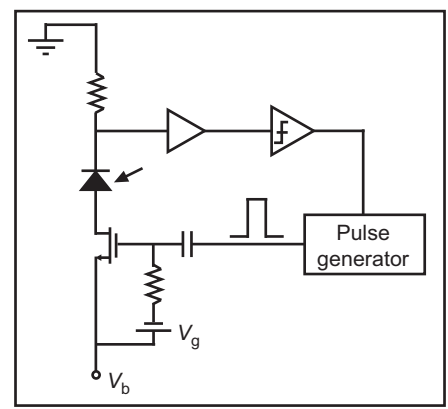

**Figure 11** Typical scheme of passive quenching and active reset for free-running SPADs. $V_b$, bias voltage; $V_g$, direct current bias to the gate terminal. SPAD, single-photon avalanche diode.

Recently a new kind of SPAD device, i.e., NFAD, was invented.[105–110] The NFAD monolithically integrates a high-resistance thin film resistor with the semiconductor structure, as shown in Figure 12a. With this design, the NFAD exploits passive quenching. Due to the integration, the parasitic capacitance of the quenching circuit is minimized so that the afterpulsing performance of the NFAD is significantly improved compared with the common SPAD. In the equivalent circuit of the NFAD (Figure 12b), integrated resistor ($R_L$) is larger than diode equivalent impedance ($R_d$) by several orders of magnitude. Therefore, avalanches can be quenched swiftly due to the voltage drop on $R_L$. When the NFAD is recharging, the recovery time is determined by $R_L C_d$. The quenching electronics of the NFAD can be similar to that of PQAR circuits, as shown in Figure 12c. Avalanches are AC coupled out. After amplification and discrimination, the outputs are driven by a hold-off time circuit. The amplified hold-off time signals are DC coupled to the anode of the NFAD. When hold-off time signals are at a high level, the bias of the NFAD is below $V_b$ due to the voltage lift at the anode of the NFAD. Using such quenching electronics, Lunghi et al.[111] demonstrated an NFAD-based detector system with a detection efficiency of 10% and a DCR of 600 Hz at −50 ℃.

In 2012, Yan et al.[112] implemented a similar free-running detector using the NFAD in which a transformer was used as an avalanche readout circuit. At −80 ℃, a detection efficiency of 10% and a DCR of 100 Hz was achieved. Using a Stirling cooler, the Geneva group significantly improved the noise performance of the free-running detector.[113] At −110 ℃, with a detection efficiency of 10%, DCR was reduced down to a record level of only 1 Hz, and $P_{ap}$ was 2.2% for 20 μs hold-off time. Such impressive performance implies that NFADs could be comparable to SNSPDs except for the limited count rates. In particular, these detectors are the ideal choice for long-distance QKD, where the limitation in count rate is not an issue.

## QUANTUM COMMUNICATION APPLICATIONS

InGaAs/InP SPADs have been widely used for optical fiber quantum communication, what we would like to illustrate in this section. Particularly in QKD, progress in transmission distance has been since ever closely linked to the progress in the development of low-noise SPDs. InGaAs/InP SPADs, in the low-frequency gating mode, were used first for QKD at the end of the 1990s using auto-compensating plug-and-play systems.[114] In 1999, Bourennane et al.[115] performed a plug-and-play experiment with a fiber transmission distance of 40 km. Hughes et al.[116] realized one-way phase-encoding QKD implementing both the BB84 and B92 protocols over 48 km of optical fiber. In 2002, Stucki et al.[117] reported a field QKD experiment over 67 km installed fiber between Geneva and Lausanne. Then Kosaka et al.[118] increased the QKD distance up to 100 km with InGaAs/InP single-photon detectors using a double-SPAD scheme as introduced previously. In 2004, the Toshiba group further increased the QKD distance up to 122 km using low DCR SPADs.[119] All the above QKD experiments used weak laser pulses (coherent states) to mimic a single-photon source. However, due to the photon-number-splitting attack, the secure distance was significantly limited. Thanks to the decoy-state scheme subsequently proposed, this primary obstacle was completely eliminated so that practical applications of QKD could be possible. In 2007, Peng et al.[120] first demonstrated a decoy-state QKD experiment over 100 km using InGaAs/InP SPADs.

Longer distances and higher rates were achieved using high-frequency gating featuring less afterpulsing. In 2007, Namekata et al.[121] first implemented a differential phase shift QKD experiment using SWG SPADs. At a clock rate of 500 MHz, the final secure key rate reached 0.33 Mbps over a distance of 15 km. The same group then implemented 24kbps secure key rate over 100 km fiber distance[122] by increasing the clock rate up to 2 GHz and optimizing the SPAD performance with an ultralow dark count probability per gate of $2.8 \times 10^{-8}$ at a detection efficiency of 6%. In 2008, Yuan et al.[123] demonstrated a GHz QKD experiment using self-differencing SPADs and achieved 27.9 kpbs secure key rate at 65.5 km. Then the Toshiba group further improved both the bit rate and the system stability to push forward GHz QKD system for practical uses.[124] In 2010, they demonstrated 1 Mbps bit rate at 50 km over a continuous operation of 36 h.[125] In 2014, they performed the coexistence experiment of GHz QKD with classical optical communication,[126] in which quantum data and bidirectional 10 Gbps classical data were combined in a single fiber using dense wavelength division multiplexing. The secure key rate reached 2.38 Mbps over 35 km and the fiber distance could be extended up to 70 km.

Due to the very short gates and hold-off times, dark count probabilities per gate are becoming pretty low and the afterpulsing may become the main noise contribution for QKD. For this reason the optimal operation temperature of SPADs can be close to room temperature, which is very convenient for commercial systems.[89,127]

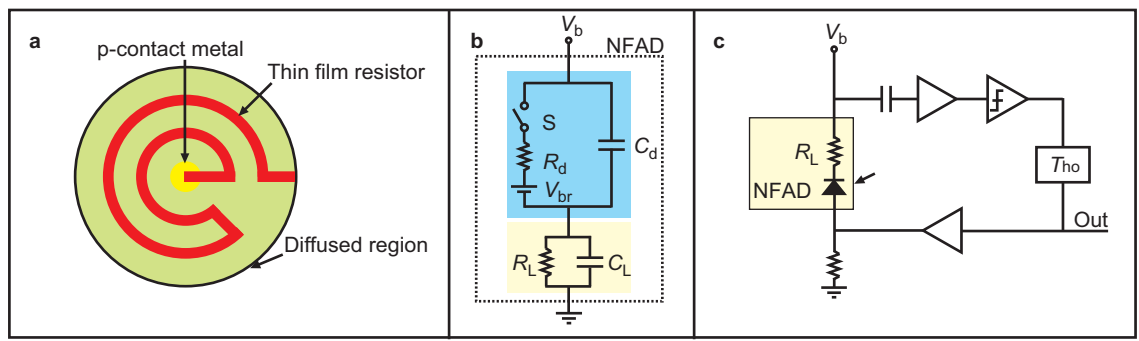

**Figure 12** Top view (**a**), equivalent circuit (**b**) and quenching electronics (**c**) of negative feedback avalanche diodes.[111] $R_d$, diode equivalent impedance. $C_d$, diode capacitance; $V_{br}$, breakdown voltage; $R_L$, integrated load resistor; $C_L$, load capacitance. $V_b$, bias voltage; $T_{ho}$, hold-off time circuit. S, switch. NFADs, negative feedback avalanche diodes.

For even longer distances, unpractical SNSPDs are generally employed. However, very recently secure QKD over more than 300 km of fiber have been realized, using ultra-low noise free-running NFADs.[128]

Apart from weak coherent states, entanglement distribution is another approach to implement QKD. In 2004, the Geneva group implemented the distribution of time-bin entangled photon pairs over 50 km fiber.[129] Takesue demonstrated experimental distribution of time-bin entanglement generated using spontaneous four-wave mixing over 60 km.[130] In 2007, the distribution of polarization entanglement over 100 km was achieved by the Vienna group.[131] Finally, Dynes et al.[132] demonstrated distribution of time-bin entangled photons over a record distance of 200 km using self-differencing InGaAs/InP SPADs.

Except QKD, other quantum communication protocols have also been implemented in optical fiber. The Geneva group demonstrated quantum teleportation over 2 km of telecom fiber[133,134] and also in an installed fiber network.[135] In 2008, Bogdanski et al.[136] reported a QSS experiment for five parties based on a single-qubit protocol. In 2013, Ma et al.[137] increased the fiber distance up to 50 km for single-qubit QSS, and circular QSS was also demonstrated in fiber.[138] Counterfactual quantum cryptography was proposed in 2009[29] and experimentally demonstrated by Ren et al.[139] and Liu et al.[140]

## CONCLUSIONS AND OUTLOOKS

III–V SPADs are the most practical tools for ultra-weak light detection in the near-infrared. In the past decades, both the academic and industrial communities have made great efforts to achieve performance improvements of SPADs. In the field of semiconductor devices, dedicated devices for single-photon detection are designed and fabricated, while APDs designed for classical optical communication are no longer widely used in photon counting applications. The device performance of the SPAD itself has been gradually improved. Also, new devices like NFADs and self-quenching SPADs[141,142] have appeared, which may particularly improve the performance of certain parameters and alleviate the requirements of quenching electronics. In the field of quenching electronics, diverse techniques have been invented for both gating and free-running operations. The gating frequency has been increased up to 2 GHz and relevant techniques have been quickly applied to QKD. In this review, we have surveyed the technical advances in low-frequency gating, high-frequency gating and free-running operation, and we have described some representative quenching schemes.

In the future, the development and evolution of SPAD-based near-infrared detectors will be continuously pushed forward in the same way. On the one hand, the SPADs themselves should have better device-level performance. This requires research and focused efforts addressing various device attributes such as structure design and optimization, high-quality material growth and fabrication technology. The key parameters to be considered in device design include PDE, DCR and $P_{ap}$. PDE and DCR are the intrinsic parameters and are generally independent from quenching electronics, which means that the parameter performance cannot be improved using different electronics at the same operating conditions. Thus, high PDE and low DCR are the two core objectives in future device design. Shrinking the size of the SPAD could be an effective approach for the reduction of DCR and $P_{ap}$, and relevant investigations have started. However, fiber coupling could be a crucial technology challenge for small-size SPADs. Another benefit of improvements in the DCR versus PDE tradeoff is operation at higher PDE given the same DCR performance. Apart from III–V SPADs, silicon-based devices could also be potential candidates for single-photon detection in the future—e.g., Ge/Si and InGaAs/Si—although currently, the DCR of these devices are still high compared with III–V SPADs.

On the other hand, quenching electronics is also important for practical detector systems. Therefore, inventing new quenching techniques and continuously optimizing the present quenching techniques are the fundamental tasks for the single-photon community. For these mature techniques, developing integrated circuits for quenching electronics is the future trend due to various advantages including cost-effectiveness, detector miniaturization, parasitic capacitance minimization and power reduction. For Si SPADs, such integrated circuits were achieved over a decade ago.[143–146] Also, integrated quenching electronics is favorable for SPAD arrays. Due to both the improvement of device performance and the integration of readout circuits, in the future, developing practical III–V SPAD arrays is possible and significant for applications requiring multi-pixel near-infrared single-photon detection.

## ACKNOWLEDGEMENTS
We acknowledge Wen-Hao Jiang for technical assistance. This work has been financially supported by the National Basic Research Program of China (Grant No. 2013CB336800), the National High-Tech R&D Program (Grant No. 2011AA010802), the National Natural Science Foundation of China (Grant No. 61275121) and the Innovative Cross-disciplinary Team Program of CAS. HZ acknowledges the financial support from the Swiss NCCR QSIT.